\documentclass[preprint,12pt]{elsarticle}

\usepackage{epsfig}

\usepackage{amssymb}

\newcounter{bla}

\journal{Computer Physics Communications}

\begin{document}

\begin{frontmatter}



\title{An upgraded issue of the parton and hadron cascade model, PACIAE 2.2}


\author[a]{Dai-Mei Zhou\corref{author}}
\author[b]{Yu-Liang Yan}
\author[b]{Xing-Long Li}
\author[b]{Xiao-Mei Li}
\author[b]{Bao-Guo Dong}
\author[a]{Xu Cai}
\author[a,b]{Ben-Hao Sa}
\cortext[author] {Corresponding author.\\\textit{E-mail address:
zhoudm@phy.ccnu.edu.cn}}
\address[a]{Key Laboratory of Quark and Lepton Physics (MOE) and
              Institute of Particle Physics, Central China Normal University,
              Wuhan 430079, China.}
\address[b]{China Institute of Atomic Energy, P. O. Box 275 (10), 102413
Beijing, China.}
\begin{abstract}
The parton and hadron cascade model PACIAE 2.1 (cf. Comput. Phys. Commun.
184 (2013) 1476) has been upgraded to the new issue of PACIAE 2.2. By this
new issue the lepton-nucleon and lepton-nucleus (inclusive) deep inelastic
scatterings can also be investigated. As an example, the PACIAE 2.2 model
is enabled to calculate the specific charged hadron multiplicity in the
$e^-$+p and $e^-$+D semi-inclusive deep-inelastic scattering at 27.6 GeV
electron beam energy. The calculated results are well comparing with the
corresponding HERMES data. Additionally, the effect of model parameters
$\alpha$ and $\beta$ in the Lund string fragmentation function on the
multiplicity is studied.
\end{abstract}

\begin{keyword}
relativistic nuclear collision; (inclusive) deep inelastic scattering,
PYTHIA model; PACIAE model.

\end{keyword}

\end{frontmatter}



{\bf PROGRAM SUMMARY}

\begin{small}
\noindent
{\em Manuscript Title: An upgraded issue of the parton and hadron cascade
model, PACIAE 2.2}                                     \\
{\em Authors: Dai-Mei Zhou, Yu-Liang Yan, Xing-Long Li, Xiao-Mei Li,
 Bao-Guo Dong, Xu Cai, and Ben-Hao Sa} \\
{\em Program Title: PACIAE version 2.2}                       \\
{\em Journal Reference:}      \\
{\em Catalogue identifier:}                                   \\
{\em Licensing provisions: none}                                   \\
{\em Programming language: FORTRAN 77 }                        \\
{\em Computer: DELL Studio XPS and/or others with a FORTRAN 77 compiler}                                               \\
{\em Operating system: Linux with FORTRAN 77 compiler}                                       \\
{\em RAM:} $\approx$ 1G bytes                                 \\
{\em Number of processors used:}                              \\
{\em Supplementary material:}                                 \\
{\em Keywords: relativistic nuclear collision; (inclusive) deep
  inelastic scattering; PYTHIA model; PACIAE model.}  \\
{\em Classification: 11.1, 17.8}                                         \\
{\em External routines/libraries:}                            \\
{\em Subprograms used:}                                       \\
{\em Catalogue identifier of previous version: }*              \\
{\em Journal reference of previous version: Ben-Hao Sa, Dai-Mei Zhou,
Yu-Liang Yan, Bao-Guo Dong, and Xu Cai, Comput. Phys. Commun.
 184(2013)1476.}\\
{\em Does the new version supersede the previous version?: Yes}   \\
{\em Nature of problem: The lepton inclusive and semi-inclusive deep
 inelastic scattering (DIS and SIDIS) off nuclear target have greatly
 contributed to the parton structure of hadron, the parametrization of
 parton distribution function (PDF), and the extraction of fragmentation
 function (FF). Unfortunately, the PACIAE 2.1 model is unable to describe
 the lepton-nucleon and lepton-nucleus DIS (SIDIS), the corresponding upgrade
 is highly required.}\\
{\em Solution method: The parton and hadron cascade model of PACIAE 2.1
 is upgraded to PACIAE 2.2 with the possibility of investigating the
 lepton-nucleon and lepton-nucleus DIS (SIDIS). In the PACIAE 2.2 model
 the lepton-nucleon and lepton-nucleus DIS are treated the same
 as proton-nucleon and proton-nucleus collisions in PACIAE 2.1,
 respectively. However, the lepton-nucleon DIS cross section is used
 instead of the nucleon-nucleon cross section in the initiation stage
 of the PACIAE model.}\\
{\em Reasons for the new version: In order that the PACIAE 2.2 model is now
 also able to simulate the lepton-nucleon and lepton-nucleus DIS.}\\
{\em Summary of revisions: In the PACIAE 2.2 model the lepton-nucleon
 and lepton-nucleus DIS are dealt with the same way as the proton-nucleon
 and proton-nucleus collisions in PACIAE 2.1, respectively. However,
 the lepton-nucleon DIS cross section is employed instead of
 nucleon-nucleon cross section.}\\
{\em Restrictions: Depend on the problem studied.}\\
{\em Unusual features: none}\\
{\em Additional comments: Email addresses: sabh@ciae.ac.cn (B.-H. Sa),
 yanyl@ciae.ac.cn (Y.-L. Yan).}\\
{\em Running time:
PACIAE 2.2 has three versions of PACIAE 2.2a, 2.2b, and 2.2c. PACIAE 2.2a
is for the elementary collision, such as pp, $\rm{\bar pp}$, and
e$^+$e$^-$ collisions, as well as the lepton-nucleon DIS, with input
file of usux.dat. PACIAE 2.2b and PACIAE 2.2c are for the nuclear-nucleus
collision of p+A and A+B as well as the lepton-nucleus DIS with input
file of usu.dat. PACIAE 2.2b and 2.2c are similar in the physical
contents but are different in the topological structure (see text for
the details).
\begin{itemize}
\item Using the attached input file of usux.dat to run 1000 events for
the $\sqrt s$=200 GeV Non Single Diffractive pp collision by PACIAE 2.2a
spends 0.5 minute.
\item Using the attached input file of usu.dat to run 10 events for the
10-40\% most central Au+Au collisions at $\sqrt{s_{NN}}$=200 GeV by
PACIAE 2.2b spends 4 minutes.
\item Using the attached input file of usu.dat to run 10 events for the
10-40\% most central Au+Au collisions at $\sqrt{s_{NN}}$=200 GeV by
PACIAE2.2c spends 8 minutes.
\end{itemize}
}
\end{small}





\section {Introduction}
The lepton inclusive and semi-inclusive deep inelastic scattering (DIS and
SIDIS) off nuclear target are the most active frontiers between
nuclear and the particle physics since the eighties of the last century.
They have greatly contributed to the parton structure of hadron \cite{jixd},
the parametrization of parton distribution function (PDF) \cite{plac,rith},
and the extraction of polarization-averaged fragmentation function (FF)
\cite{lead,herm1}. They also play important role in the hadronization of
initial partonic state and the space-time evolution of the fragmentation
process \cite{herm1}.

Two new electron ion collider (EIC) programs of the eRHIC at BNL and ELIC
at Jefferson Laboratory (JLab) are evolved in the USA \cite{usa}. They are
aimed at reaching the highly variable center of mass (cms) energies of $\sim$
20  - 150 GeV and the high collision luminosity of $\sim 10^{33-34}$
cm$^{-2}$s$^{-1}$. Meanwhile, a similar program of LHeC is also progressed
at CERN in Europe \cite{euro}. The collision cms energy and luminosity of
LHeC may achieve $\sim$1 TeV and $\sim 10^{33}$cm$^{-2}$s$^{-1}$, respectively
. Both the eRHIC (ELIC) and LHeC are able to yield great insight into
the nucleon structure, such as how partons share the spin, mass, and magnetic
moment of a nucleon, etc. To confront this expected new DIS era, an
upgraded PACIAE 2.2 model, being able to investigate the l-p and l-A DIS, is
introduced.
\begin{figure}[]
\begin{center}
\includegraphics[width=0.3\textwidth]{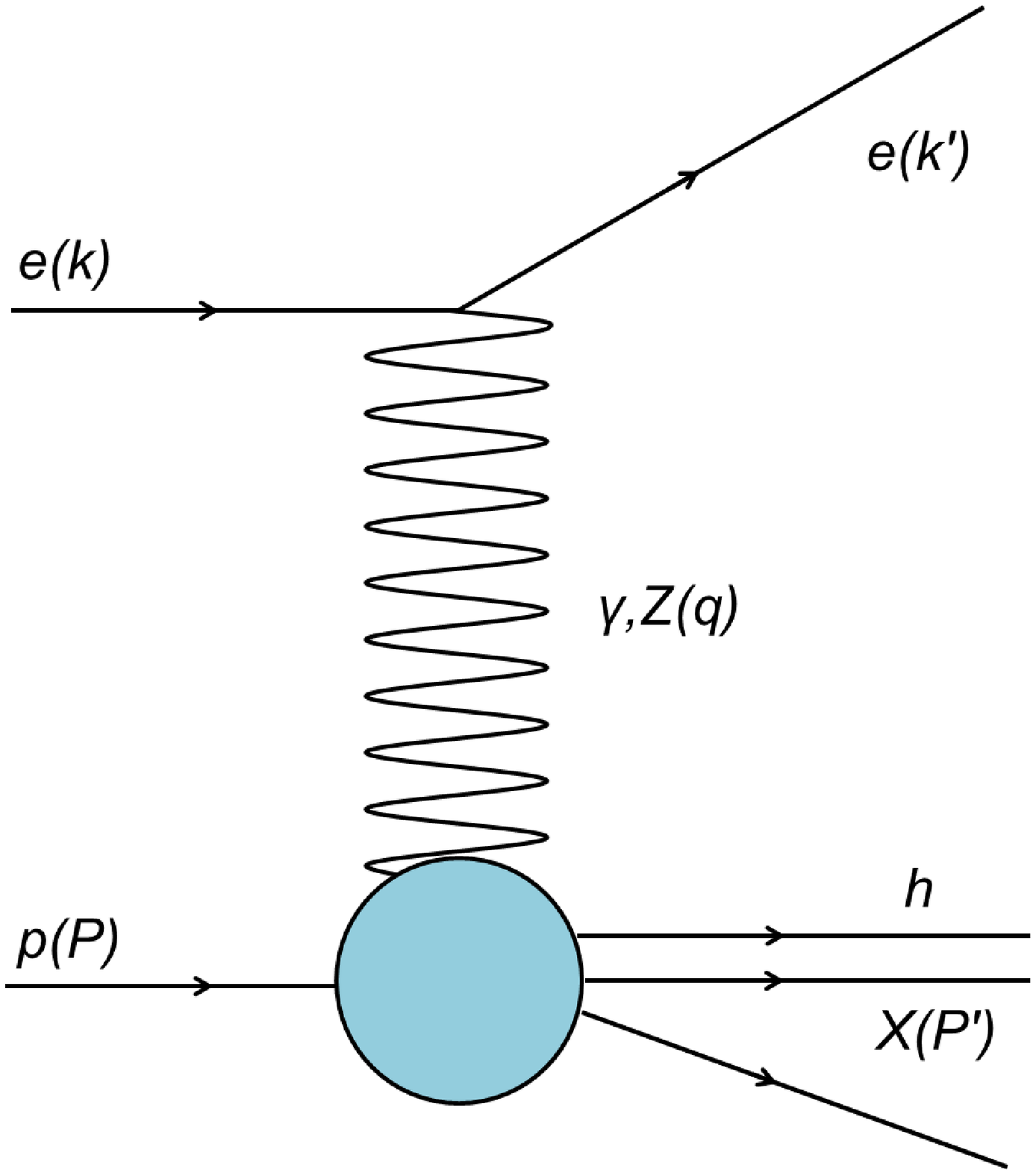} \hspace{2.5cm}
\includegraphics[width=0.3\textwidth]{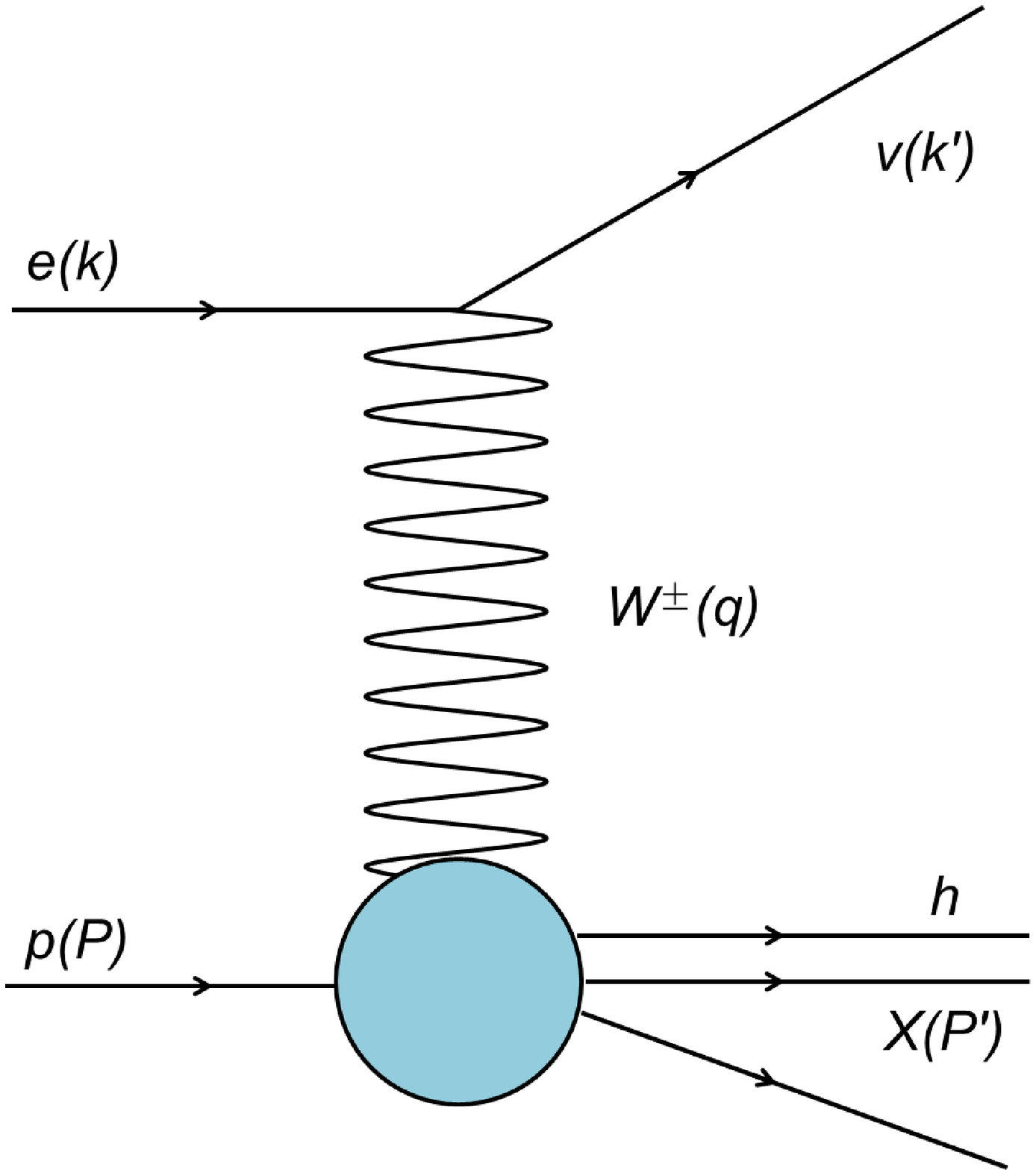}
\caption{(color online) Lowest order (Born approximation) Feynman diagram
of the neutral current (NC) and charged current (CC) electron DIS.}
\label{nccc}
\end{center}
\end{figure}

For a reliable extraction of the FF with the quark fragmentation function
of $D_q^h$ distinguished from the antiquark ones of $D_{\bar q}^h$, the
data of specific charged hadron ($\pi^+$, $\pi^-$, $K^+$, $K^-$)
multiplicity in the unpolarized SIDIS are highly needed \cite{lead,herm2}.
Recently, the HERMES collaboration has measured the multiplicity of
charged pions and kaons in the 27.6 GeV electron beam SIDIS off proton
and deuteron \cite{herm2}. As an example, the PACIAE 2.2 model is employed
calculating the DIS normalized $\pi^+$, $\pi^-$, $K^+$, and $K^-$
multiplicities in the above electron SIDIS off proton and the deuteron.
The calculated results are well comparing with the corresponding HERMES
data \cite{herm2}.

We have given a summary for the history of the PACIAE model in introduction
section of the first long writing-up paper of PACIAE 2.0 \cite{sa1}. Since
then, the most important progresses are:
\begin{itemize}
\item The net-proton, net-baryon, and net-charge multiplicity nonstatistical
fluctuations in the relativistic nuclear collisions are successfully
investigated in \cite{zhou1,zhou2}.
\item We have updated the PACIAE model from PACIAE 2.0 to PACIAE 2.1
\cite{sa2} by randomly sampling the $p_x$ and $p_y$ components of hadrons,
generated in the string fragmentation, on a circumference of ellipse with
half major and minor axes of $p_T(1+\delta_p)$ and $p_T(1-\delta_p)$
instead of on a circle with radius $p_T$ originally. The PACIAE 2.1 model
has been successfully employed studying systematically the elliptic flow
parameter in the relativistic nuclear collisions at RHIC and LHC energies
\cite{sa3}.
\end{itemize}

The paper is organized as follows: After the introduction of section I
the PACIAE model and its new content of the lepton-nucleon DIS
are briefly presented in section II. Here it is distinguished that the
PACIAE 2.2 model has three versions of PACIAE 2.2a, 2.2b, and 2.2c. For
the elementary collisions of pp, $\rm{\bar pp}$, and e$^+$e$^-$ as well
as the lepton-nucleon DIS, PACIAE 2.2a should be used together with the
input file of usux.dat. PACIAE 2.2b and PACIAE 2.2c are used for the
nuclear-nucleus collisions of p+A and A+B as well as the lepton-nucleus
DIS with the input file of usu.dat. PACIAE 2.2b is similar to PACIAE 2.2c
in the physical contents but is different in the topological structure
(see later for the details). The results are given in section III. In
the section IV a brief conclusions are drawn.
\section {Models}
The PACIAE model \cite{sa1,sa2} is based on PYTHIA \cite{sjos}. However,
the PYTHIA model is for high energy elementary collisions ($e^+e^-$,
lepton-hadron, and hadron-hadron ($hh$) collisions) only, but PACIAE is
also for lepton-nucleus DIS and nuclear-nucleus collisions (p+A and A+B).
In the PYTHIA model a $hh$ collision, for instance, is decomposed into
parton-parton collisions. The hard parton-parton collision is described
by the LO-pQCD parton-parton interactions with the modification of parton
distribution function in a hadron. The soft parton-parton collision, a
non-perturbative process, is considered empirically. The initial- and
final-state QCD radiations as well as the multiparton interactions are
taken into account. So the consequence of a $hh$ collision is a
partonic multijet state composed of diquarks (anti-diquarks),
quarks (antiquarks), and gluons, besides a few hadronic remnants.
It is followed by the string construction and fragmentation, thus a
final hadronic state is obtained for a $hh$ ($pp$) collision finally.
\begin{figure}[]
\begin{center}
\includegraphics[width=1.\textwidth]{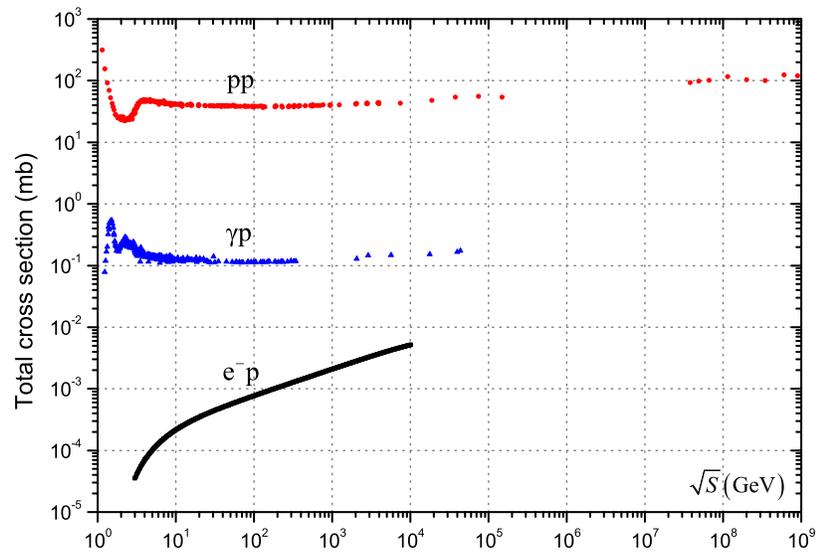}
\caption{(color online) pp, $\gamma$p, and $e^-$p (DID) total cross sections.}
\label{tcros}
\end{center}
\end{figure}

In the PACIAE model \cite{sa1,sa2}, the nucleons in a colliding nucleus
are first randomly distributed according to the Woods-Saxon distribution
in the spatial coordinate space. The participant nucleons, resulted from
Glauber model calculation, are required to be inside the overlap zone,
formed when two colliding nuclei path through each other at a given
impact parameter. The spectator nucleons are required to be outside
the overlap zone but inside the nucleus-nucleus collision system.
If the incident beam is in the $z$ direction, we set $p_x=p_y=0$ and
$p_z=p_{beam}$ for the projectile nucleons, $p_x=p_y=p_z=0$ for the target
nucleons in the laboratory framework as well as $p_x=p_y=0$ and $p_z=
-p_{beam}$ for the target nucleons in the collider framework.

We then decompose a nucleus-nucleus collision into nucleon-nucleon
($NN$) collisions according to nucleon straight-line trajectories and
the $NN$ total cross section. Each $NN$ collision is dealt by PYTHIA
with the string fragmentation switched-off and the diquarks
(anti-diquarks) broken into quark pairs (anti-quark pairs). A partonic
initial state (composed of the quarks, antiquarks, gluons, and a few
hadronic remnants) is obtained for a nucleus-nucleus collision after
all of the $NN$ collision pairs were exhausted.

This partonic initial stage is followed by a parton evolution stage.
In this stage the parton rescattering is performed by the Monte Carlo
method with $2\rightarrow2$ LO-pQCD cross sections \cite{comb}.
The hadronization stage follows the parton evolution stage. The Lund
string fragmentation model and a phenomenological coalescence model are
provided for the hadronization. However, the string fragmentation model
is selected in these calculations. The rescattering among produced hadrons is
then dealt with the usual two body collision model \cite{sa1,sa2}. In this
hadronic evolution stage, only the rescatterings among $\pi$, $K$, $\rho
(\omega)$, $\phi$, $p$, $n$, $\Delta$, $\Lambda$, $\Sigma$, $\Xi$, $\Omega$,
and their antiparticles are considered for simplicity.

As the same as PACIAE 2.0 and 2.1, PACIAE 2.2 also has three versions of
PACIAE 2.2a, 2.2b, and 2.2c. PACIAE 2.2a describes the relativistic
elementary collisions of pp, $\rm{\bar pp}$, e$^+$e$^-$, and lepton-nucleon
DIS with input file of usux.dat. PACIAE 2.2b and PACIAE 2.2c describe the
relativistic nuclear-nucleus collisions of p+A and A+B as well as
lepton-nucleus DIS. In the PACIAE 2.2b model the partonic initiation,
partonic evolution (rescattering), hadronization, and the hadronic evolution
(rescattering) are performed for each hh collision pairs independently until
all the hh collision pairs are collided. Oppositely, in the PACIAE 2.2c, the
partonic initiation is first performed for all the hh collision pairs. This
full initial partonic state is proceeded to the partonic evolution stage,
then the hadronization stage, and at last the hadronic evolution stage.
Therefore, PACIAE 2.2b and 2.2c are similar in the physical contents but are
different in the topological structure.

For p+p and p+A (A+p) collisions, the overlap zone is not introduced
presently. We deal the l+p and l+A DIS like the p+p and p+A collisions,
respectively, but instead of the nucleon-nucleon cross section the
lepton-nucleon DIS cross section is used. In the PYTHIA model, the $e^-$+p
($e^+$+p ), $\mu^-$+p ($\mu^+$+p ), and $\tau^-$+p ($\tau^+$+p ) DIS
have two options of '$e^-$' ('$e^+$'), '$\mu^-$' ('$\mu^+$'), and '$\tau^-$'
('$\tau^+$') as well as '$gamma/e^-$' ($gamma/e^+$), '$gamma/\mu^-$'
('$gamma/\mu^+$'), and '$gamma/\tau^-$' ('gamma/$\tau^+$'),
respectively, in the specification of beam and target particle. However,
the neutrino-nucleon (antineutrino-nucleon) DIS has only the first option.
In order to be more consistent and to have a better running stability the
first option is chosen for all the lepton-nucleon DIS in the PACIAE 2.2
model.

Fig.~\ref{nccc} shows the leading order (Born approximation) Feynman
diagrams for the neutral current (NC, the exchange of $\gamma/Z$ boson,
left panel) and charged current (CC, the exchange of $W^\pm$ boson,
right panel) $e^-$+p DIS. There are two vertices in the left panel of
Fig.~\ref{nccc}, for instance. At the upper boson vertex the initial
state QED and weak radiations have to be considered. At the lower boson
vertex, not only the leading order parton level process of
$V^*q\rightarrow q$ ($V^*$ refers to $\gamma/Z/W$) but also the first order
QCD radiation of $V^*g\rightarrow qg$ as well as the boson-gluon fusion
process of $V^*g\rightarrow q\bar q$ have to be considered. Furthermore, the
parton shower approach has been introduced to take higher than first order
QCD effects into account \cite{inge}. Therefore the DIS cross section can be
formally expressed as
\begin{equation}
\sigma_{NC(CC)}=\sigma_{NC(CC)}^{Born}(1+\delta_{NC(CC)}^{qed})
                (1+\delta_{NC(CC)}^{weak})(1+\delta_{NC(CC)}^{qcd})
\end{equation}
\cite{adlo}, where $\sigma_{NC(CC)}^{Born}$ is the Born cross section,
$\delta_{NC(CC)}^{qed}$, $\delta_{NC(CC)}^{weak}$, and $\delta_{NC(CC)}
^{qcd}$ are, respectively, the QED, weak, and the QCD radiative corrections.
\begin{figure}[]
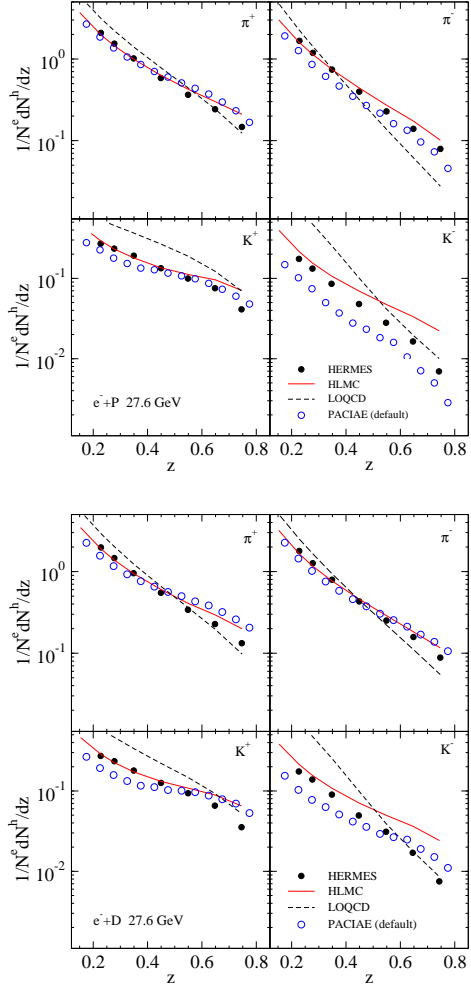

\begin{center}
\includegraphics[width=0.45\textwidth]{ep.eps}\\
\vspace{0.5cm}
\includegraphics[width=0.45\textwidth]{ed.eps}
\caption{(color online) Multiplicity of DIS normalized specific charged
hadron as a function of $z$ in the $e^-+$p (upper panel) and $e^-+$D (lower
panel) DIS at 27.6 GeV electron beam energy.}
\label{epd0}
\end{center}
\end{figure}

In the lowest-order perturbative QCD theory, the NC/CC DIS Born cross
section of the unpolarized electron on an unpolarized nucleon can be
expressed by the structure functions $F_1, F_2, F_3$ as follows
\cite{pdg}
\begin{equation}\label{epDIS}
\frac{{\rm{d^2}} \sigma^I}{{\rm{d}} x {\rm{d}} y}=
\frac{4\pi \alpha^2}{xyQ^2}\eta^I\left( \left(1-y-\frac{x^2y^2M^2}{Q^2}
\right)F_2^I+y^2xF_1^I\mp \left(y-\frac{y^2}{2}\right)xF_3^I\right),
\end{equation}
where the mass of the incident and scattered leptons are neglected. In
the above equation, $I$ denotes $NC$ or $CC$. $\alpha$ stands for the
fine structure constant. $x\equiv x_B$ and $y$ are the Bjorken scaling
variable and fraction energy of $\gamma/Z/W$ boson, respectively. $Q^2$
is the negative squared 4-momentum transfer. $M$ refers to the mass of target
nucleon. $\eta^{NC}=1$, $\eta^{CC}=(1\pm\lambda)^2\eta_W$, and
\begin{eqnarray}
\eta_W=\frac{1}{2}\left(\frac{G_FM_W^2}{4\pi\alpha}\frac{Q^2}
 {Q^2+M_W^2}\right) \\
G_F=\frac{e^2}{4\sqrt{2}sin^2\theta_WM_W^2},
\end{eqnarray}
where $M_W$ and $\theta_W$ are the mass of $W$ boson and the Weinberg angle,
respectively. $\lambda =\pm 1$ is the helicity of the incident lepton.

The structure functions above can be expressed by the parton
distribution function of nucleon in the quark-parton model \cite{bjor}.
Presently we can not calculate PDF from the first principles and PDF can
only be extracted from the QCD fits with a measure of the agreement between
the experimental data of lepton-nucleon DIS cross sections and the
theoretical models \cite{gizh}. With the PDFs at hand, we are able to
calculate the lepton-nucleon DIS cross section. The black curve in
Fig.~\ref{tcros} gives the unpolarized $e^-$+p DIS total cross section
calculated with HERAPDF1.5 LO PDF set \cite{coop}. In the calculation
\cite{li} the cuts are first set for $Q^2>1$ GeV and $W^2>1.96$ Gev
($W^2$ is the squared invariant mass of the photon-nucleon system) and then
the cuts in $x$ and $y$ are derived according to the relationships among
kinematic variables and $cos^2\theta\leq 1$. The red and blue data points in
Fig.~\ref{tcros} are, respectively, the total cross section of pp and
$\gamma$p collisions copied from \cite{pdg}.

One knows well that the incident proton, in the p+Au collisions at RHIC
energies for instance, may collide with a few ($\sim$2-5) nucleons when
it passes through the gold target. Since the $e^-$+p DIS total cross
section is a few order of magnitude smaller than the pp collision at the
range of $\sqrt s<$1000 GeV (cf. Fig.~\ref{tcros}), one may expect
that the incident electron, in this energy range, may suffer at most one
DIS with the nucleon when it passes through the target nucleus. The
struck nucleon is the one with lowest approaching distance from the
incident electron. This is the same for other incident leptons because
the lepton-nucleon DIS total cross section is not so much different among
the different leptons \cite{li}.

Therefore in the pioneer studies \cite{gisin} for the lepton-nucleus DIS by
PYTHIA + BUU transport model, the FRITIOF 7.02 \cite{pi} or PYTHIA 6.2
\cite{sjos2} was employed to generate a lepton-nucleon DIS event. The
generated hadronic final state was then input into the BUU
(Boltzmann-Uehling-Uhlenbeck) equation \cite{buu} considering the final state
hadronic interaction (hadronic rescattering). This PYTHIA + BUU transport
model successfully described the HERMES data of the ratio of DIS normalized
charged hadron multiplicity in the lepton-A (nucleus) DIS to the one in the
lepton-deuteron DIS, for the 27.5 GeV electron beam energy $e^++^{14}$N and
$e^++^{84}$Kr DIS \cite{herm3}.
\section {Results}
As mentioned in \cite{herm2,hill} the measured hadron multiplicity in
SIDIS has first to be corrected for the radiative effects, the limitations in
geometric acceptances, and the detector resolution. The Born-level
multiplicity is then obtained. Therefore, the DIS (total yield) normalized
Born-level multiplicity of the $h$ type hadrons as a function of $z$ (the
fractional energy of hadron $h$) in the lepton SIDIS off a nuclear target
can be expressed as
\begin{equation}
\frac{1}{N_{DIS}}\frac{dN^h}{dz}=
                  \frac{1}{N_{DIS}}\int d^5N^h(x_B,Q^2,z,P_{h\bot},
                  \phi_h)dx_BdQ^2dP_{h\bot}d\phi_h
\label{multi}
\end{equation}
\cite{herm2}, where $N_{DIS}$ refers to the DIS total yield, $P_{h\bot}$ is the
component of the hadron momentum $P_h$ transverse to $q$ (the 4-momentum of
the mediator $\gamma/Z/W$) and $\phi_h$ stands for the azimuthal angle
between the lepton scattering plane and the hadron production plane. Thus one
may compare the calculated $\frac{1}{N_{DIS}}\frac{dN^h}{dz}$ in the full
kinematic phase space with the HERMES data, like done in \cite{herm2,hill}.
\begin{figure}[]
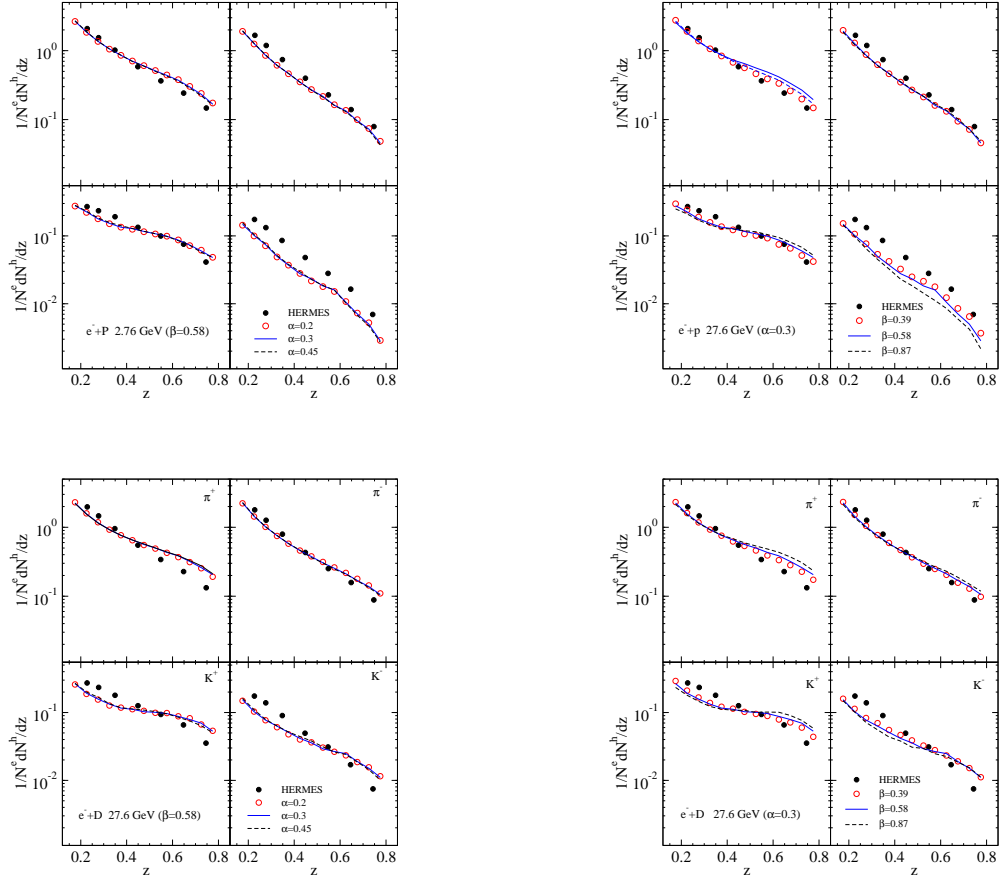

\begin{center}
\includegraphics[width=0.38\textwidth]{ep2.eps} \hspace{2.5cm}
\includegraphics[width=0.38\textwidth]{ep1.eps} \\ \vspace{1.0cm}
\includegraphics[width=0.38\textwidth]{ed2.eps} \hspace{2.5cm}
\includegraphics[width=0.38\textwidth]{ed1.eps}
\caption{(color online) The effect of parameter $\alpha$ (left panels)
         and $\beta$ (right panels) in the Lund string fragmentation
         function on $\frac{1}{N_{DIS}}\frac{dN^h}{dz}$ in the $e^-+$p
         (upper panels) and $e^-+$D (lower panels) DIS at 27.6 GeV electron
         beam energy.}
\label{epd12}
\end{center}
\end{figure}

In the PACIAE 2.2 model simulations, the model parameters are all fixed
the same as the default values given in PYTHIA. Figure \ref{epd0} gives the
comparison of the HERMES $\frac{1}{N_{DIS}}\frac{dN^h}{dz}$ data (solid
circles) \cite{herm2} to the corresponding results of the PACIAE 2.2 model
(open circles), HLMC (solid line), and the LOQCD (dashed line). Here HLMC
refers to the HERMES Lund Monte Carlo. HLMC is a combination of the DIS
event generator Lepto \cite{inge} based on JETSET 7.4 and PYTHIA 5.7
\cite{sjos1}, the detector simulation program based on GEANT \cite{geant},
and the HERMES reconstruction program \cite{hill}. The HLMC results in
Fig.~\ref{epd0} were calculated with fitting thirteen model parameters to the
multiplicity as a function of $z$, $p_T$ (transverse momentum), and $\eta$
(pseudorapidity) of the $\pi^-$, $K^-$, and $\bar p$ \cite{hill}. The LOQCD
results in Fig.~\ref{epd0} were calculated in the framework of collinear
factorization at the leading order perturbative QCD \cite{herm2,kret}. In
Fig.~\ref{epd0} one sees that the default PACIAE fairly well reproduces
the HERMES data.
\begin{table*}[htbp]
\tabcolsep 2.mm
\centering \caption{DIS normalized multiplicity of $\pi^+$, $\pi^-$, $K^+$,
           and $K^-$ in the $e^-+$p DIS.}
\begin{tabular}{cccccccccc}
\hline\hline
\multicolumn{5}{c}{Effect of $\alpha$ with default $\beta$=0.58} &
\multicolumn{5}{c}{Effect of $\beta$ with default $\alpha$=0.3}\\
$\alpha$& $\pi^+$& $\pi^-$& $K^+$& $K^-$& $\beta$& $\pi^+$& $\pi^-$& $K^+$& $K^-$\\
0.2 &1.331 &0.9591 &0.1164 &0.04406 &0.39 &1.355 &0.9834 &0.1204 &0.04828 \\
0.3 &1.339 &0.9666 &0.1177 &0.04646 &0.58 &1.339 &0.9666 &0.1177 &0.04646 \\
0.45&1.351 &0.9785 &0.1196 &0.04732 &0.87 &1.315 &0.9419 &0.1132 &0.04161 \\
\hline
\hline
\end{tabular}
\label{epd12_1}
\end{table*}
\begin{table*}[htbp]
\tabcolsep 2.mm
\centering \caption{DID normalized multiplicity of $\pi^+$, $\pi^-$, $K^+$,
           and $K^-$ in the $e^-+$D DIS.}
\begin{tabular}{cccccccccc}
\hline\hline
\multicolumn{5}{c}{Effect of $\alpha$ with default $\beta$=0.58} &
\multicolumn{5}{c}{Effect of $\beta$ with default $\alpha$=0.3}\\
$\alpha$& $\pi^+$& $\pi^-$& $K^+$& $K^-$& $\beta$& $\pi^+$& $\pi^-$& $K^+$& $K^-$\\
0.2 &1.314 &1.364 &0.1203 &0.05538 &0.39 &1.351 &1.407 &0.1269 &0.06116 \\
0.3 &1.326 &1.383 &0.1219 &0.05696 &0.58 &1.326 &1.383 &0.1219 &0.05696 \\
0.45&1.345 &1.399 &0.1239 &0.05919 &0.87 &1.288 &1.342 &0.1161 &0.05192 \\
\hline
\hline
\end{tabular}
\label{epd12_2}
\end{table*}

Meanwhile, the effect of $\alpha$ and $\beta$ parameters in the Lund
string fragmentation function \cite{sjos}
\begin{equation}
f(\hat z)\propto \frac{1}{\hat z}(1-\hat z)^{\alpha}
                 \exp({-\frac{\beta m_T^2}{\hat z}})
\end{equation}
on the multiplicity is investigated. In the above equation $\hat z$
refers to the fraction lightcone variable taken by the fragmented
hadron out of the fragmenting particle and $m_T^2=p_T^2+m_0^2$ where $m_0$
refers to the rest mass of the fragmented hadron. The results are given in
the tables \ref{epd12_1} and \ref{epd12_2} as well as in figure
\ref{epd12}. We see here that the multiplicity increases (decreases) with
 $\alpha$ ($\beta$) increasing. The effect shown in the differential
observable $\frac{1}{N_{DIS}}\frac{dN^h}{dz}$ is weak but visible, as shown
in  Figs~\ref{epd12}.

\section {Conclusions}
In summary, we have upgraded the the parton and hadron cascade model
PACIAE 2.1 to a new issue of PACIAE 2.2 involving the lepton-nucleon and
lepton-nucleus DIS. The PACIAE 2.2 is then employed investigating the DIS
normalized specific charged hadron multiplicity as function of $z$,
$\frac{1}{N_{DIS}}\frac{dN^h}{dz}$, in the 27.6 GeV electron beam energy
$e^-+$p and $e^-+$D SIDIS. The PACIAE 2.2 results calculated with
default parameters reproduce fairly well the corresponding HERMES data
\cite{herm2}.

Additionally, we have investigated the effect of model parameters $\alpha$
and $\beta$ in the Lund string fragmentation function. It turned out that
the particle multiplicity increases (decreases) with $\alpha$ ($\beta$)
increasing. The effect on the global observable of yield is not small, but
on the differential observable of $\frac{1}{N_{DIS}}\frac{dN^h}{dz}$ is
just visible. These effects are expected to be increased with
increasing reaction energy and system size.

Acknowledgments: This work was supported by the National Natural Science
Foundation of China under grant Nos.:11075217, 11105227, 11175070,
11477130 and by the No. 111 project of the foreign expert bureau of China.
BHS would like to thank C. P. Yuan for HERAFitter, G. Schnell for HERMES
data, and Y. Mao and S. Joosten for discussions. YLY acknowledges the
financial support from SUT-NRU project under contract No. 17/2555.


\end{document}